\documentclass[12pt]{article}
\usepackage{epsf}
\usepackage{citesort,psfrag}

\renewcommand{\thefootnote}{\#\arabic{footnote}}
\begin{document}

\newcommand{\gtrsim}{ \mathop{}_{\textstyle \sim}^{\textstyle >} }
\newcommand{\lesssim}{ \mathop{}_{\textstyle \sim}^{\textstyle <} }

\renewcommand{\thefootnote}{\fnsymbol{footnote}}
\setcounter{footnote}{0}
\begin{titlepage}

\def\thefootnote{\fnsymbol{footnote}}

\begin{center}

\hfill RESCEU-17/01\\
\hfill TU-628\\
\hfill astro-ph/0108081\\
\vskip .5in

{\Large \bf
Isocurvature Fluctuations \\
in Tracker Quintessence Models
}

\vskip .45in

{\large
Masahiro Kawasaki$^{(a)}$,
Takeo Moroi$^{(b)}$ and Tomo Takahashi$^{(b)}$
}

\vskip .45in

{\em
$^{(a)}$Research Center for the Early Universe, School of Science,
University of Tokyo\\
Tokyo 113-0033, Japan
}
\vskip .2in

{\em
$^{(b)}$Department of Physics,  Tohoku University\\
Sendai 980-8578, Japan
}

\end{center}

\vskip .4in

\begin{abstract}

We consider effects of the isocurvature perturbation in the framework
of the tracker-type quintessence models.  During the inflation,
fluctuations in the amplitude of the quintessence field are generated,
which provide isocurvature component of the cosmic density
perturbation.  Contrary to the conventional notion, it is shown that
effects of the isocurvature fluctuation may become sizable in some
cases, and in particular, the cosmic microwave background angular
power spectrum may be significantly enhanced due to the effect of the
isocurvature mode.  Such an enhancement may be detectable in 
on-going and future experiments.

\end{abstract}

\end{titlepage}

\renewcommand{\thepage}{\arabic{page}}
\setcounter{page}{1}
\renewcommand{\thefootnote}{\#\arabic{footnote}}
\setcounter{footnote}{0}

Recent cosmological observations suggest that there exists a dark
energy which must be added to the matter density in order to reach the
critical density \cite{krauss_2001}. Although the cosmological
constant is usually assumed as the dark energy, in the past years, a
slowly evolving scalar field, dubbed as ``quintessence'' has been
proposed as the dark energy \cite{caldwell_et_al_1998}. There are some
differences between the cosmological constant and
quintessence. Firstly, for the quintessence, the equation-of-state
parameter $\omega_Q = \rho_Q / p_Q $ varies with time, whilst for the
cosmological constant, it remains a fixed value $\omega_{\Lambda} = -
1$.  Secondly, since the quintessence is a scalar field, it
fluctuates.

One of the observational discriminator of these models is the cosmic
microwave background (CMB) anisotropies. Many authors have studied the
effects of the quintessence field on the CMB anisotropy in various
models. As mentioned above, the quintessence field fluctuates, so its
fluctuations should be taken into account in calculating the CMB
anisotropy.  Generally, the initial fluctuations of the quintessence
fields are generated during inflation. Since these fluctuations behave
as the isocurvature mode, there can exist both adiabatic and
isocurvature perturbations.  Implications of the isocurvature modes
have been studied in \cite{aph0101014}, and in particular for the
cosine-type quintessence model, it was pointed out that the
isocurvature fluctuations may be detectable in on-going and future
experiments by the present authors \cite{kawasaki_et_al_2001}.

In this letter, we focus on the effects of the isocurvature
fluctuations on the CMB anisotropy in the tracker-type quintessence
models. The tracker-type quintessence models have attractor-like
solutions which alleviate the initial condition problems. When we take
initial conditions such that the quintessence field starts to roll
down the potential at early epoch, damping effect on fluctuation of
the quintessence field is significant. So, in this case, even if we
take a nonzero value of primordial quintessence-field fluctuation, the
fluctuation goes to zero because of the damping effects.  But we will
show that, in some cases, the damping effect is not so significant. In
such cases, the effect of the isocurvature fluctuations may be sizable
and affect the CMB angular power spectrum.

Since the effects of the isocurvature fluctuations depend on models of
quintessence, first, we classify the tracker-type quintessence models.
Although many models have been proposed
\cite{aph0101014,skordis_albrecht_2000,ferreira_joyce_1998,sahni_wang_2000,urena_lopes_matos_2000,brax_martin_2000,brax_martin_1999,dedelson_et_al_2000,caldwell_et_al_1998,brax_et_al_2000,ratra_peebles_1988,coble_et_al_1997,viana_liddle_1998,waga_friemann_2000,ferreira_joyce_1997,balbi_et_al_2000,doran_et_al_2001,barreiro_et_al_2000},
they can be classified into two groups, using the evolution of the
energy density of the quintessence $\rho_Q$. The evolution of the
energy density of the tracker-type models can be written in a simple
form.  Tracker-type quintessence models, as previously mentioned, have
an attractor-like solution in a sense that a very wide range of
initial conditions converge to a common evolutionary track. We call
the epoch when this attractor-like solution is realized as ``tracking
regime.''  In the tracking regime, $\omega_Q$ is almost constant and
$1-\omega_Q^2$ is significantly different from zero
\cite{steinhard_et_al_1999}. This means that the kinetic and potential
energy of the quintessence have a fixed ratio. Thus, the relation
$\dot{Q}^2 \propto V(Q) $ holds in the tracking regime.  When this
relation is satisfied, we can show that the energy density of the
quintessence is proportional to $a^{-n}$, where $n$ is a constant. On
the contrary, the energy density of the dominant component of the
universe is written as $\rho_D \propto a^{-m}$ (for the matter and
radiation dominations, $m=3$ and 4, respectively).  When $n=m$, the
energy density of the quintessence can be a significant fraction of
the total energy density of the universe from earlier epoch.  On the
contrary, when $n<m$, $\rho_Q$ cannot be significant at early
epoch.\footnote{
The case with $n>m$ is also possible \cite{Liddle_Scherrer_1998}, but
almost all model which have been proposed so far, can be categorized
into the type-I or the type-II models. So we will not consider the case 
with $n >m$ in this letter.  }
For a purpose of this letter, we will consider two models as
representative models; ``AS model'' \cite{skordis_albrecht_2000} and
``Ratra-Peebles model'' \cite{ratra_peebles_1988} for models with
$n=m$ and $n<m$, respectively.

Before we show the evolution of the quintessence energy density from
numerical calculations, we investigate it analytically. To study the
evolution of the quintessence fields, we write down the basic
equations.  In a spatially flat homogeneous universe which contains a
perfect fluid with energy $\rho$ and pressure $p$ and a quintessence
field $Q$ with energy $\rho_{Q}$ and pressure $p_{Q}$, the Friedmann
equation becomes
\begin{eqnarray}
    H^2 = \frac{1}{3 M_*^2} ( \rho + \rho_{Q} ),
    \label{eq:friedmann}
\end{eqnarray}
where $H$ is the Hubble parameter and $M_* \simeq 2.4 \times 10^{18}$
GeV is the reduced Planck scale.  The energy density and pressure of
the quintessence field are
\begin{eqnarray}
  \label{eq:quintsns_energy_press}
  \rho_{Q} = \frac{1}{2} \dot{Q}^{2} + V(Q), ~~~~~ p_{Q} = \frac{1}{2}
  \dot{Q}^{2} - V(Q),
\end{eqnarray}
where the dot represents the derivative with respect to time $t$.  The
quintessence field $Q$ obeys the equation of motion
\begin{eqnarray}
  \label{eq:quintsns_eq_of_motion}
  \ddot{Q} + 3 H \dot{Q} + \frac{dV}{dQ} = 0.
\end{eqnarray}
The energy conservation equation for each component is, in case where
there is no energy exchange
\begin{eqnarray}
  \label{eq:energy_cons}
  \dot{\rho} = -3 H (\rho + p).
\end{eqnarray}
Thus, when the universe is dominated by a perfect fluid with the energy
density $\rho_D \propto a^{-m}$, the scale factor behaves as
\begin{eqnarray}
  \label{eq:a_time_dep}
  a \propto t^{2/m}.
\end{eqnarray}
In this case, from Eqs.(\ref{eq:quintsns_energy_press}) and
(\ref{eq:quintsns_eq_of_motion}),
\begin{eqnarray}
  \label{eq:potential_der}
  \ddot{Q} +\frac{6}{mt} \dot{Q} + \frac{dV}{dQ}=0, 
\end{eqnarray}

First, let us discuss the evolution of $\rho_Q$ for the AS model which
has the potential of the form
\begin{eqnarray}
   \label{eq:AS_potential}
  V(Q) = \left[
            (Q -b)^{2} + a \right] e^{-\lambda Q} = f(Q) e^{-\lambda Q} ,
\end{eqnarray}
where $a,b$ and $\lambda$ are model parameters.  Since $f(Q)$ changes
more slowly than $e^{-\lambda Q}$ in the tracking regime, in the
analytic investigations below, we approximately treat $f(Q)$ as a
constant.  Substituting Eq.(\ref{eq:AS_potential}) into
Eq.(\ref{eq:potential_der}), we obtain
\begin{eqnarray}
  \label{eq:as_EoM}
  \ddot{Q} + \frac{6}{mt} \dot{Q} - f(Q) \lambda e^{-\lambda Q} =0.
\end{eqnarray}
The solution to this equation is $ Q=A \ln \lambda Bt$, where
$A$ and $B$ are constants.  In the tracking regime,
$\dot{Q}^2$ is proportional to $V(Q)$, thus $\rho_Q \propto t^{-2} $
in this model. Using Eq.(\ref{eq:a_time_dep}), it is shown that
\begin{eqnarray}
  \label{eq:as_rho}
  \rho_Q \propto a^{-m}.
\end{eqnarray}
Therefore, in the AS model, the energy density of the quintessence has
the same $a$ dependence as that of the dominant component of the
universe. This means that the energy density of the quintessence can
be a sizable fraction of the total energy density of the universe at
early epoch.  In fact, the above result does not depend on the
detailed structure of the function $f(Q)$ as far as $f(Q)$ is a slowly
varying function on the tracking regime.  Thus, the same discussion is
applicable to other types of quintessence models with
$V(Q)=f(Q)e^{-\lambda Q}$.

Next, we consider the Ratra-Peebles model which has the potential of
the form,
\begin{eqnarray}
    \label{eq:RP_potential}
    V(Q) = \frac{\Lambda^{4 + \alpha}}{Q^{\alpha}},
\end{eqnarray}
where $\Lambda$ and $\alpha ( >0) $ are model parameters. With this
potential, the equation of motion of the quintessence field becomes
\begin{eqnarray}
    \label{eq:RP_EoM}
    \ddot{Q} + \frac{6}{mt} \dot{Q} - \alpha 
    \frac{\Lambda^{4 + \alpha}}{Q^{\alpha +1}} =0.
\end{eqnarray}
This equation has the solution $Q = C t^{\nu}$ where $\nu =
2/(2+\alpha)$ and $C$ is a constant. Since $\alpha$ is taken to be
positive, the value of $\nu$ is $ 0 < \nu < 1$. The energy density of
the quintessence in this model becomes
\begin{eqnarray}
    \label{eq:RP_rho}
    \rho_Q \propto a^{-m(1-  \nu)}.
\end{eqnarray}
Thus, in the Ratra-Peebles model, $\rho_Q$ decreases more slowly than
that of the dominant component of the universe, which implies that the
energy density of the quintessence cannot be a significant fraction of
the energy density in the early universe.

\begin{figure}[t]
 \begin{center}
    \scalebox{1}{\hspace{-2cm}\includegraphics{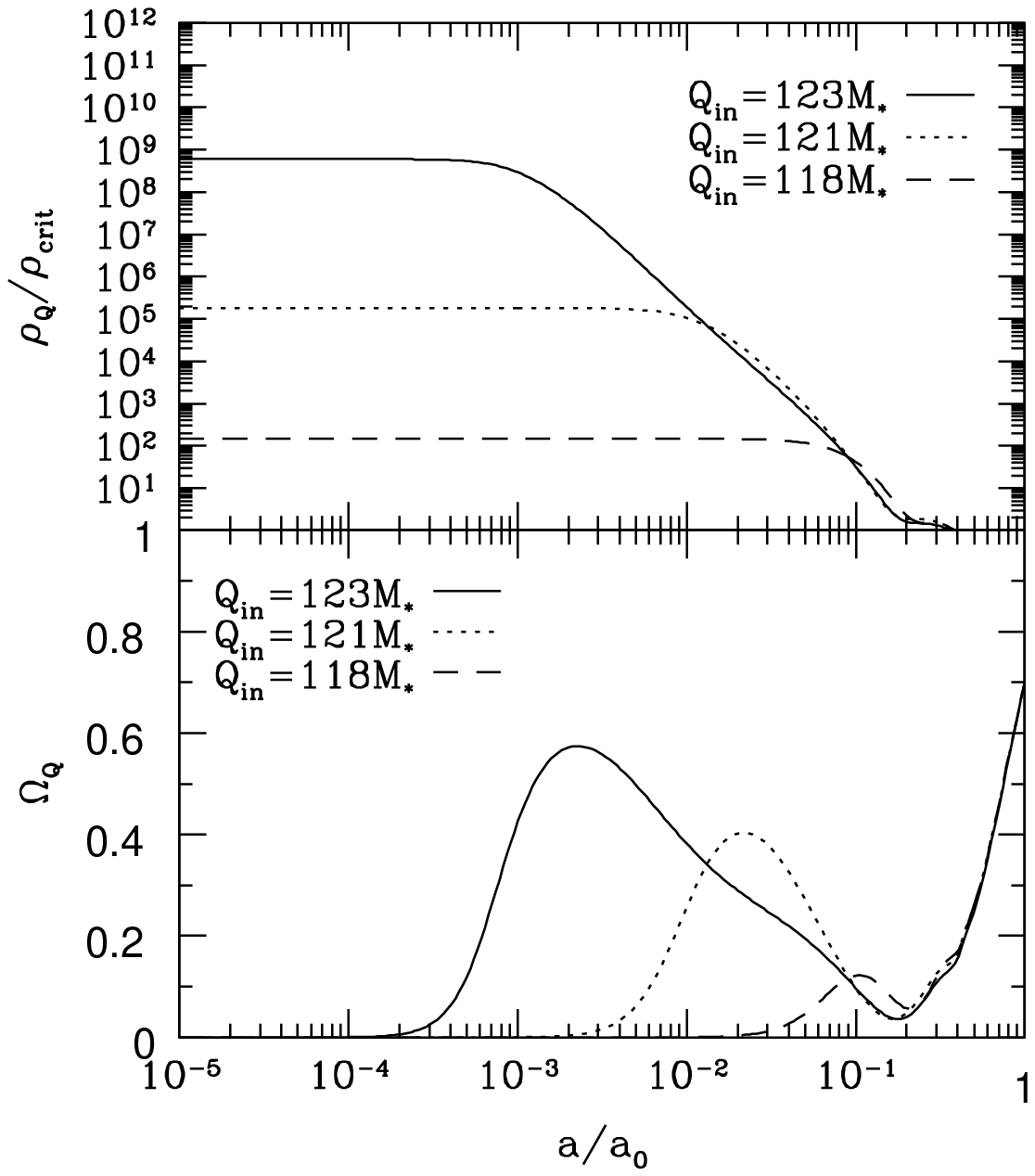}}
  \end{center}  
  \vspace{-5cm}
    \caption{Evolution of the energy density in the AS model. 
    The model parameters are taken to be $\lambda = 2.2/M_*, A=0.01
    M_*^2$ and $B=123.6805 M_* $.  The initial amplitude are $Q_{\rm
    in}=123 M_*$ (solid line), $121 M_*$ (dotted line), and $118
    M_*$ (dashed line).  The cosmological parameters are taken
    to be $\Omega_0 = 0.3$, $\Omega_b h^2 = 0.019$ and $h=0.65$, where
    $h$ is the Hubble parameter in units of 100 km/s/Mpc.}
    \label{fig:density_as}
\end{figure}

Now we show evolutions of the energy density of the quintessence
models from the numerical calculations.  In Fig.\ref{fig:density_as},
we show the evolution of the quintessence energy density for the AS
model, where we take several values of the initial amplitude for the
quintessence field.  From the figure, we can read off when the
quintessence field starts to enter the tracking regime for several
values of the initial amplitudes.  We can also see the property of the
AS model; the energy density of the quintessence field can have
$O(0.1)$ contributions to the total energy density of the universe
from early epoch.

\begin{figure}[t]
 \begin{center}
    \scalebox{1}{\hspace{-2cm}\includegraphics{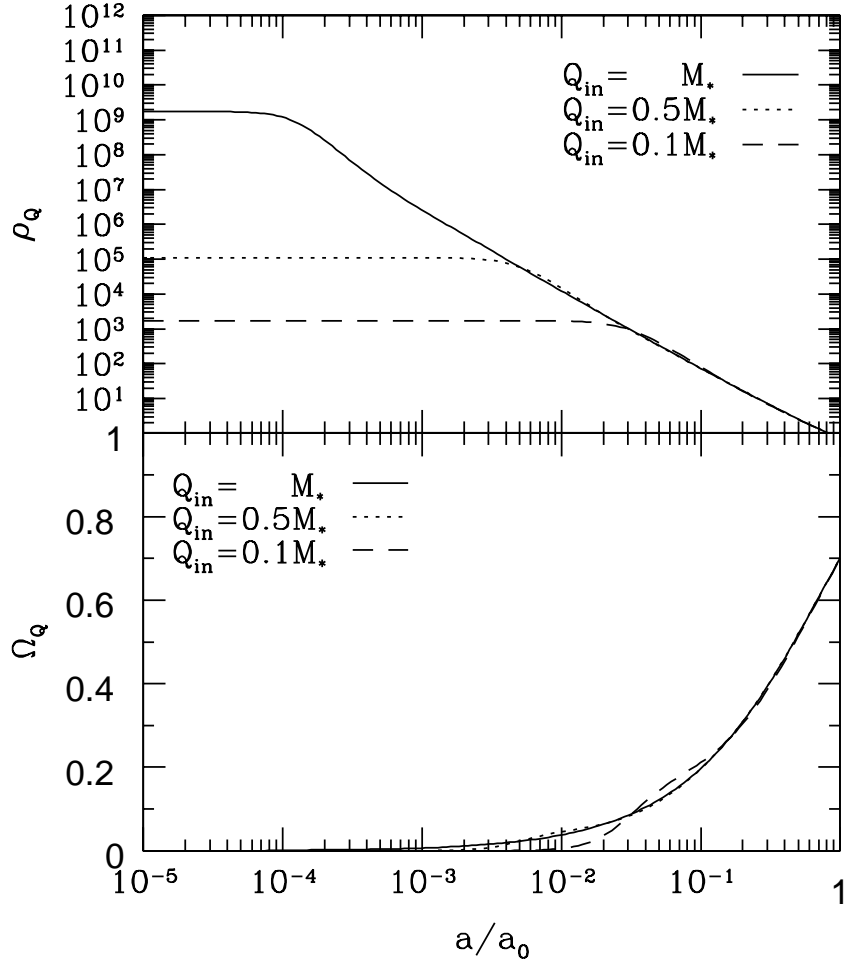}}
  \end{center}  
  \vspace{-5cm}
    \caption{Evolution of the energy density in the Ratra-Peebles model.
    The parameter we take here, $\Lambda = 5.122 \times 10^{6}$ GeV
    and $\alpha = 6$. The initial amplitudes of the quintessence field
    are $Q_{\rm in} =M_*$ (solid line), $0.5M_*$ (dotted line)
    $0.1M_*$ (dashed line). The
    cosmological parameters are the same as those in Fig.\ref{fig:density_as}.}
    \label{fig:density_ratra_peebles}
\end{figure}

In Fig.\ \ref{fig:density_ratra_peebles}, the evolution of the energy
density are shown for the Ratra-Peebles model. Different from the AS
model, the quintessence energy density of this type of model is
negligible for $ z \gg O(1)$.  This fact has an important implication
to the effects of the isocurvature fluctuations on the CMB angular
power spectrum.

Next let us discuss the effects of the isocurvature fluctuations on
the CMB angular power spectrum.  For this purpose, we study the
evolutions of the fluctuations of the quintessence analytically. We
decompose the quintessence field $Q$ as
\begin{eqnarray}
  Q (t,\vec{x}) = \bar{Q} (t) + q (t, \vec{x}),
\end{eqnarray}
where $q$ is the perturbation of the amplitude of the quintessence
field.
The equation of motion for $q$ is, in the conformal Newtonian gauge,
\begin{eqnarray}
    \ddot{q} + 3 H \dot{q} - \left( \frac{a}{a_0} \right)^{-2} 
    \partial_i^2 q
    + \frac{d^2V(\bar{Q})}{dQ^2} q = 
(\dot{\Psi} -3 \dot{\Phi})\bar{Q} 
-2 \left(\frac{a}{a_0} \right)^2 \frac{dV}{dQ} \Psi ,
    \label{eq:eq_of_motion_dq}
\end{eqnarray}
where the perturbed line element in the conformal Newtonian gauge is
given by
\begin{eqnarray}
  ds^2 &=& - (1 +2 \Psi) dt^2 
+ \left( \frac{a}{a_0} \right)^2 (1 + 2 \Phi) \delta_{ij} 
  dx^idx^j  \nonumber \\ 
&=&
 \left( \frac{a}{a_0} \right)^2 
\left[ - (1 +2 \Psi) d\tau^2  +
(1 + 2 \Phi) \delta_{ij} dx^idx^j \right],
\end{eqnarray}
where $a$ is the scale factor at time $t$, $a_0$ the scale factor at
the present time and $\tau$ is the conformal time.

From Eq.(\ref{eq:eq_of_motion_dq}), we can show the damping behavior
of the quintessence field in the tracking regime.  To solve this
equation, we write down the second derivative of the potential as a
function of the sound speed $c_s^2$ of a quintessence field $Q$
\cite{brax_et_al_2000}
\begin{eqnarray}
  \frac{d^2V(\bar{Q})}{dQ^2} = 
  \frac{3}{2} H \dot{c_s^2} + \frac{3}{2} H^2 ( c_s^2 -1 )
  \left[ \frac{\dot{H}}{H^2} - \frac{3}{2} ( c_s^2 + 1 ) \right],
  \label{eq:2nd_der_V}
\end{eqnarray}
where the sound speed of the quintessence field is written as
\begin{eqnarray}
  c_s^2 = \frac{\dot{p_Q}}{\dot{\rho_Q}} =
 \frac{\ddot{\bar{Q}} - dV/dQ}{\ddot{\bar{Q}} + dV/dQ} .
\end{eqnarray}
Since the kinetic energy is proportional to the potential energy in
the tracking regime, the relation $\ddot{Q} \propto dV/dQ$ holds.  So
the sound speed of the quintessence field
$c_s^2$ is a constant during this regime.  Therefore $\dot{c_s^2}$ can
be set to zero in Eq.(\ref{eq:2nd_der_V}).  When the quintessence is a
subdominant component of the universe, the Hubble parameter $H$ can be
written using the equation-of-state parameter $\omega_D = p_D /
\rho_D$ of the dominant component of the universe,
\begin{eqnarray}
    H = \frac{2}{3(1+\omega_D )t}.
    \label{eq:hubble2time}
\end{eqnarray}
To study the isocurvature mode at the superhorizon scale, we 
neglect the $k$ dependence and the right hand side of
Eq.(\ref{eq:eq_of_motion_dq}).  Then, using the above relations, the
equation of motions can be written as
\begin{eqnarray}
    \ddot{q} + \frac{2}{ (1+\omega_D)t } \dot{q} 
    + \frac{1}{(1+\omega_D)^2 t^2 }(1-c_s^2)(c_s^2 + \omega_D +2 ) 
    q =0.
\end{eqnarray}
This equation has power-law solutions like $q \propto t^{\xi}$, where
the power-law index $\xi$ is
\begin{eqnarray}
    \xi = \frac{\omega_D -1}{2(\omega_D +1)} \left[ 1 \pm \sqrt{ 1 -
        \frac{4}{(\omega_D -1)^2}(1-c_s^2)(c_s^2 + \omega_D +2 ) }
    \right].
    \label{eq:power_index}
\end{eqnarray}
Since $c_s^2$ is equal to $\omega_Q$ in the tracking regime
\cite{brax_et_al_2000} and $\omega_{D,Q}$ cannot be larger than unity,
the real part of $\xi$ is always negative.  Therefore $q$ damps with
time.  It follows that if we take $q \ne 0$ initially, $q$ damps to
zero in case that the quintessence field experiences long period of
tracking \cite{brax_et_al_2000,aph0101014}.  But, in the case that the
quintessence field enters the tracking regime at later time, the
fluctuations may not damp so much. In this case, the isocurvature
fluctuations (i.e., $q \ne 0$) can affect CMB anisotropies.

The primordial fluctuations in the quintessence amplitude is generated 
in the early universe, probably during the inflation.
When the mass of the quintessence field is negligible compared to 
Hubble parameter during inflation $H_{\rm inf}$,
the primordial fluctuation is \cite{linde_1990}\footnote{
If $m_{q} \gtrsim H_{\rm inf}$, the primordial fluctuation damps to
zero rapidly and cannot affect the CMB anisotropies. So we do not
consider such a case.
}
\begin{eqnarray}
    q_{\rm inf}(k) = \frac{H_{\rm inf}}{2 \pi}.
    \label{eq:initial_amp}
\end{eqnarray}

To parameterize the size of the isocurvature contribution, we define
the following quantity $r_q$:
\begin{eqnarray}
    r_q \equiv
    \frac{q_{\rm RD}}{M_*\Psi^{\rm (adi)}_{\rm RD}},
    \label{eq:q/Psi}
\end{eqnarray}
where $q_{\rm RD}$ is the fluctuation of the quintessence field in the
deep radiation-dominated epoch when the quintessence field is
slow-rolling, while $\Psi^{\rm (adi)}_{\rm RD}$ is $\Psi$ from the
adiabatic mode (i.e., contribution from the inflaton-field
fluctuation) in the deep radiation-dominated epoch.  For a given model
of slow-roll inflation, $\Psi^{\rm (adi)}_{\rm RD}$ is
\cite{bardeen_et_al_1983}
\begin{eqnarray}
    \Psi^{\rm (adi)}_{\rm RD} = \frac{4}{9}
    \left( \frac{H_{\rm inf}}{2\pi} 
        \frac{V_{\rm inf}}{M_*^2V_{\sf inf}'} 
    \right),
\end{eqnarray}
where $V_{\rm inf}$ is the inflaton potential and $V'_{\rm inf}$ is
its derivative with respect to the inflaton field $\chi$.  

The parameter $r_q$ depends on the detailed scenario of cosmology.  In
a simple case with $q_{\rm RD}=q_{\rm inf}$,
\begin{eqnarray}
    r_q \simeq \frac{9}{4} \frac{M_*V'_{\rm inf}}{V_{\rm inf}}.
\end{eqnarray}
For example, for the chaotic inflation with $V_{\rm inf}\propto\chi^p$
where $p$ is an integer, $r_q$ is given by
\begin{eqnarray}
    r_q|_{\rm chaotic} = \frac{9p M_*}{4 \chi (k_{\rm COBE})},
\end{eqnarray}
where $\chi (k_{\rm COBE})$ represents the inflaton amplitude at the
time when the COBE scale crosses the horizon.  Numerically, we found
$r_q\simeq 0.3$ $-$ 0.6 for $p=2$ $-$ 10.  The ratio $r_q$ may become
larger, however, in more complicated cases.  For example, if the
kinetic function for the quintessence field $Z_Q$ varies during
inflation, $q$ also changes.  This may happen if $Z_Q$ depends on the
inflaton field.  Therefore, in general the ratio $r_q$ is model
dependent and we treat $r_q$ as a free parameter in our analysis.

Now we show the results from numerical calculations. We used the
modified version of CMBFAST \cite{cmbfast} to calculate the CMB
angular power spectrum $C_l$ which is defined as
\begin{eqnarray}
    \left\langle \Delta T(\vec{x},\vec{\gamma}) 
        \Delta T(\vec{x},\vec{\gamma}') \right\rangle 
    = \frac{1}{4\pi} \sum_l
    (2l+1) C_l P_l (\vec{\gamma} \cdot \vec{\gamma}'),
\end{eqnarray}
where $\Delta T(\vec{x},\vec{\gamma})$ is the temperature fluctuation
of the CMB pointing to the direction $\vec{\gamma}$, and $P_l$ is the
Legendre polynomial.  The average is over the position $\vec{x}$.

\begin{figure}[htbp]
  \begin{center}
    \scalebox{0.85}{\includegraphics{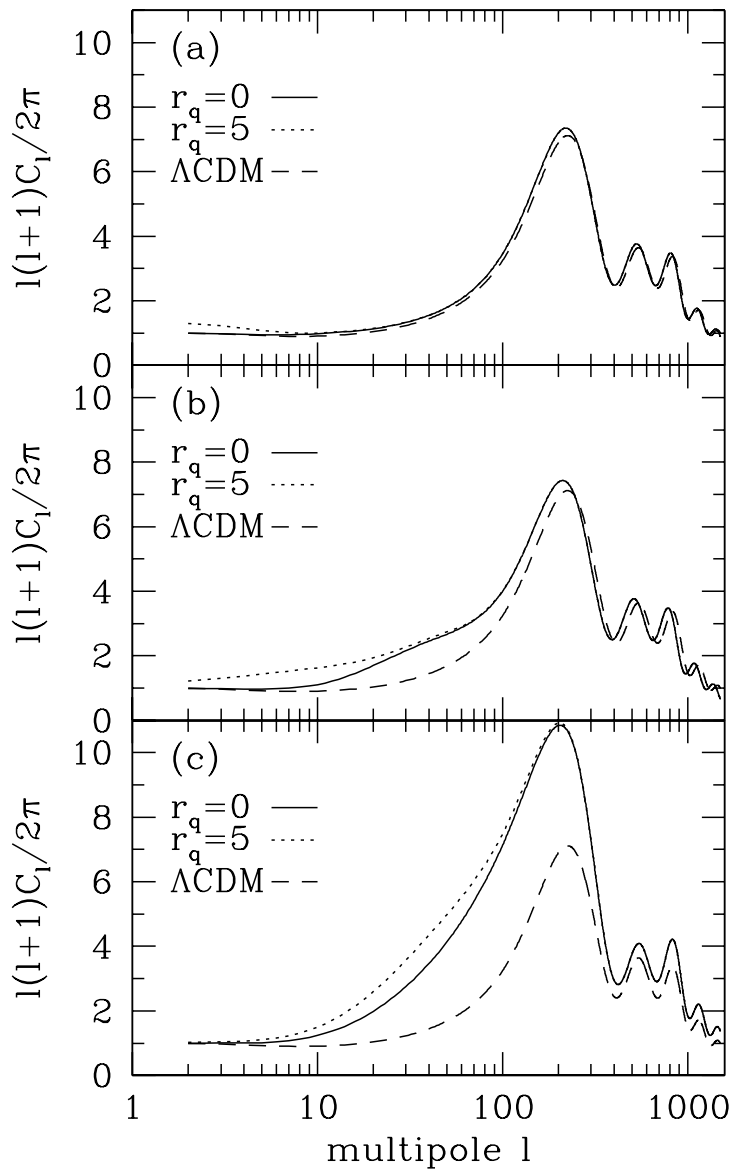}}
  \end{center}
    \caption{The CMB angular power spectrum in the AS model. 
      The initial amplitudes are (a) $Q_{\rm in} =123M_*$, (b) $Q_{\rm
        in} =121M_*$, and (c) $Q_{\rm in} =118 M_*$.  The cosmological
      and model parameters are the same as those in Fig.\ 
      \ref{fig:density_as}. The overall normalization is arbitrary in
      this figure.}
    \label{fig:cmb_iso_as}
\end{figure}

In Fig.\ \ref{fig:cmb_iso_as}, we plot the CMB angular power spectrum
$C_l$ in the AS model with several values of the initial amplitude
$Q_{\rm in}$ and the ratio $r_q$.\footnote{
  With the cosmological and model parameters used in Fig.\ 
  \ref{fig:cmb_iso_as}, we checked that theoretical predictions of
  $C_l$ are in good agreements with current observational data with
  proper normalization.  }
For comparison, we also plotted $C_l$ in the cosmological constant
case.

First, we discuss the adiabatic case (i.e., $r_q=0$), in particular,
focusing on differences between the quintessence and the cosmological
constant cases.  One can see that the locations of the acoustic peaks
may differ in two cases.  In our case, locations of the peaks mostly
depend on the angular diameter distance to the last scattering surface
(LSS) $r_{\theta}(\tau_{*})$, and the location of the $n$-th peak in
the $l$ space is estimated as \cite{Hu_Sugiyama_1995,Hu_Sugiyama_1996}
\begin{eqnarray}
    l_{n} \simeq \frac{r_\theta (\tau_*)}{r_{s}(\tau_{*})} n \pi,
\end{eqnarray}
where $r_{s}(\tau_{*})$ is the sound horizon at the recombination. The
angular diameter distance becomes smaller when the energy density of
the universe is large since the expansion rate becomes larger in this
case.  The energy density of the universe in the earlier epoch becomes
larger in the case with the quintessence. Thus, in the quintessence
case, the locations of the acoustic peaks are shifted to lower
multipoles due to this effect.  Such a shift is observed in Fig.\ 
\ref{fig:cmb_iso_as}.

Another point we should mention is the height of acoustic peaks
(relative to $C_l$ at lower multipoles).  In the AS model, the driving
effect is important to understand the height of the peaks; in the
radiation-dominated epoch, the amplitudes of the acoustic oscillation
is boosted just after the horizon crossing due to the decay of $\Psi$.
Compared to the cosmological constant case, the expansion rate of the
universe becomes larger in the quintessence case, resulting in faster
decay of the gravitational potential $\Psi$.  This effect enhances
the driving effect.  Thus, in particular in the AS models, height of
the acoustic peaks are enhanced relative to lower multipoles
\cite{skordis_albrecht_2000}.

Next we consider the case with $r_q\ne 0$.  For this purpose, it is
instructive to study the behavior of the perturbations generated by
the fluctuation of the quintessence amplitude.  Evolutions of the
perturbations are governed by the following equations; for the metric
perturbation,
\begin{eqnarray}
    k^2 \Phi = 4 \pi G \left(\frac{a}{a_0} \right)^2 \rho_{\rm tot}
    \left[ \delta_{\rm tot} + 3 \mathcal{H} ( 1 + \omega_{\rm tot} )
        \frac{V_{\rm tot}}{k} \right],
\end{eqnarray}
and for the photon, at the superhorizon scale,
\begin{eqnarray}
    \delta'_\gamma = -\frac{3}{4} k V_\gamma - 4 \Phi',~~~
    V'_\gamma = k \left( \frac{1}{4}\delta_\gamma + \Psi \right),
\end{eqnarray}
where ${\cal H}\equiv a'/a$, $\delta_\gamma$ and $V_\gamma$
($\delta_{\rm tot}$ and $V_{\rm tot}$) are density perturbation and
velocity perturbation of photon (those of total matter density) which
are defined in the Newtonian gauge, and the prime represents the
derivative with respect to the conformal time $\tau$.  In the deep
radiation-dominated epoch when the quintessence field is slow-rolling,
the fluctuations in energy density, pressure and velocity of the
quintessence are approximated as
\begin{eqnarray}
    \delta \rho_Q \simeq -\delta p_Q \simeq 
    \frac{dV}{dQ} q_{\rm RD}, ~~~
    (\rho_Q + p_Q) V_Q = k \left(\frac{a}{a_0} \right)^{-2} Q' 
    q_{\rm RD}.
\end{eqnarray}
In the deep radiation-dominated epoch when the scale we concern is out
of horizon, we can study the behavior of the perturbations by
expanding each variables with the conformal time $\tau$.  Neglecting
the shear stress from neutrino (i.e., using $\Pi_{\rm tot}=0$),
$\Psi\simeq - \Phi$. Thus we find
\begin{eqnarray}
    \Psi \simeq C q_{\rm RD}\tau^4,~~~
    \delta_\gamma \simeq 4 C q_{\rm RD}\tau^4, ~~~
    V_\gamma \simeq \frac{2}{5} Ck q_{\rm RD}\tau^5,
    \label{Psi(tau)}
\end{eqnarray}
with
\begin{eqnarray}
    C= - \frac{1}{126} \frac{dV}{dQ} \frac{1}{M_*^4} 
    \rho_{\rm rad0},
\end{eqnarray}
where $\rho_{\rm rad0}$ is the energy density of radiation at the
present epoch. Here, we keep the leading term in $\tau$. In the
radiation-dominated epoch, $\tau\propto t^{1/2}$ and all of these
variables (i.e., $\Psi$, $\delta_\gamma$ and $V_\gamma$) grow with
time for the isocurvature mode.  In addition, we can see that, at the
superhorizon scale, $\Psi$ has the same $k$ dependence as $q_{\rm
  RD}$. Thus, if $q_{\rm RD}$ has no scale dependence, $\Psi$ is
scale-invariant.

\begin{figure}[t]
  \begin{center}
    \scalebox{0.85}{\hspace{0cm}\includegraphics{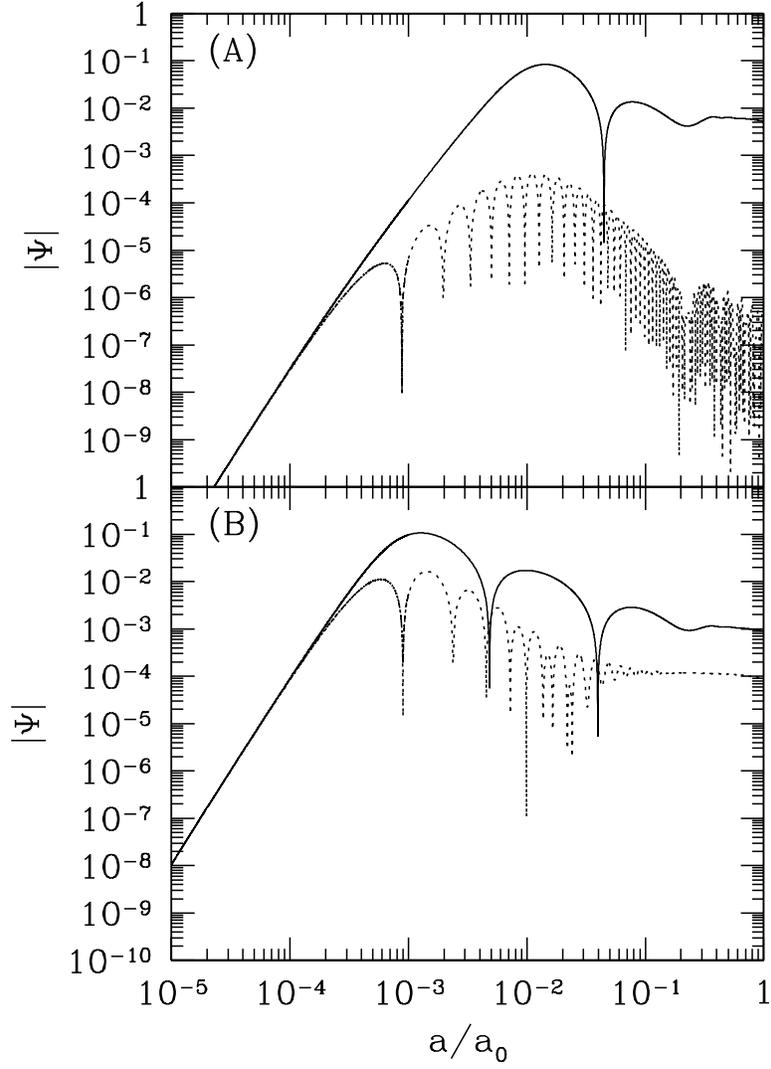}}
  \end{center}  
  \vspace{-2cm}
    \caption{The evolutions of $|\Psi|$ of purely isocurvature mode
      with (A) $Q_{\rm in} =121 M_*$  and (B) $Q_{\rm in}=118 M_*$.  We
      take $ k=1 \times 10^{-4} h ~{\rm Mpc}^{-1}$ (solid line) and $
      k=1 \times 10^{-2} h ~{\rm Mpc}^{-1}$ (dotted  line).  The model
      and cosmological parameters are the same as those in Fig.\ 
      \ref{fig:density_as}.}
    \label{fig:Psi_AS}
\end{figure}

This behavior can be confirmed by numerical calculations.  In Fig.\ 
\ref{fig:Psi_AS}, we show evolution of the gravitational potential
generated by the fluctuation in the quintessence amplitude.  When the
quintessence is in the slow-roll regime, $\Psi$ is proportional to
$a^4$.  (Notice that, in the radiation-dominated epoch,
$a\propto\tau$.)  At the time when the quintessence field enters the
tracking regime, $\Psi$ ceases to grow since $q$ starts to damp.
Thus, $\Psi$ has its maximum value at this epoch.  Consequently, the
CMB angular power spectrum is the most enhanced at the angular scale
which enters the horizon at the time when the tracking regime starts.

In the AS model, since $\Omega_Q$ can be $O(0.1)$ in the tracking
regime, fluctuation of the quintessence field may significantly affect
the angular power spectrum $C_l$.  To see the effect, in Fig.\ 
\ref{fig:cmb_iso_as}, we plot the CMB angular power spectrum for the
case with $r_q=5$.  (Notice that the enhancement due to the
isocurvature contribution, $C_l(r_q \ne 0)/C_l(r_q=0)-1$, is
proportional to $r_q^2$.)  As we can see from the figure, when $r_q
\ne 0$, the angular power spectrum $C_l$ is enhanced.

\begin{figure}[htbp]
    \begin{center}
        \scalebox{1}{\hspace{-1.5cm}
        \includegraphics{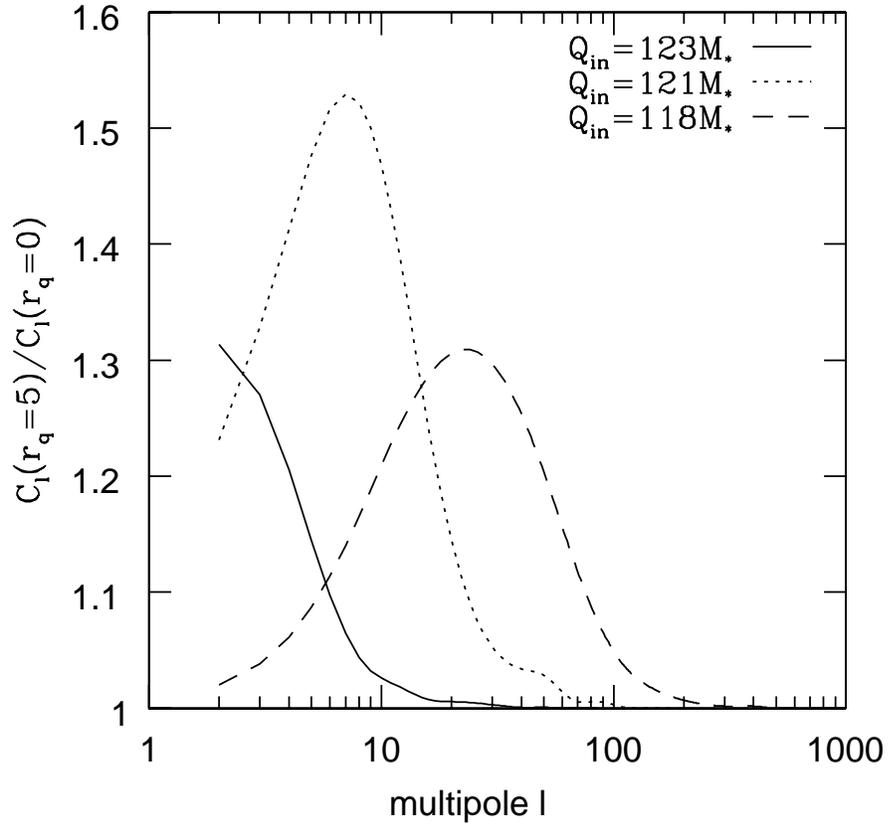}}
    \end{center}
    \vspace{-4cm}
    \caption{The ratio $C_l (r_q=5)/C_l (r_q=0)$ in the AS model.
      The initial amplitude are taken to be $123 M_*$ (solid line),
      $121 M_*$ (dotted line), and $118 M_*$ (dashed line).
      The cosmological and model parameters are the same as those in Fig.\ 
      \ref{fig:density_as}. Notice that the quantity $C_l(r_q
      \ne 0) / C_l(r_q=0) -1 $ is proportional to $r_q^2$. }
    \label{fig:as020110_cmb_iso_adi_ratio}
\end{figure}

To study the enhancement in $C_l$ by the isocurvature fluctuation, we
plot the ratio $C_l(r_q=5)/C_l(r_q=0)$ with several values of the
initial amplitude in Fig.\ \ref{fig:as020110_cmb_iso_adi_ratio}. We
can clearly see that the multipole with the largest enhancement
depends on the initial amplitudes of the quintessence fields.  As we
discussed before, this angular scale is determined by the time when
the tracking regime starts; the effect of the isocurvature fluctuation
is not significant for the angular scale much larger than and much
smaller than this scale because of the damping effect of $q$ or
smallness of $\Omega_Q$.  As we can see, the later the quintessence
field enters the tracking regime, $C_l$ is most enhanced at lower
multipole $l$.  Since the epoch when the quintessence field enters the
tracking regime depends on its initial amplitude, the enhancement in
$C_l$ may give information about the initial condition of the
quintessence fields.  In addition, with $r_q=5$, the enhancement in
$C_l$ can be as large as 50 \% relative to the case with $r_q=0$ when
$Q_{\rm in}=121 M_*$, which can be detectable in on-going and future
satellite experiments \cite{MAP,PLANCK}.  Thus studying effects of
isocurvature fluctuations provides us implications on model building of
the quintessence, which is complementary to studying the adiabatic
fluctuation.

Finally, we comment on the case with the Ratra-Peebles model.  We also
calculated the CMB angular power spectrum in the Ratra-Peebles model.
However, in this model, effects of isocurvature fluctuation are very
small since $\Omega_Q$ is very small in the early universe.  For
example, even if we take $r_q = 5$ with the model parameters $\Lambda =
5.122 \times 10^{6}$ GeV and $\alpha = 6$, enhancement in $C_l$ is
less than a few \%.

{\sl Acknowledgment:} The authors would like to thank an anonymous
referee for the crucial comment on the scale dependence of the metric
perturbation. This work is supported by the Grant-in-Aid for
Scientific Research from the Ministry of Education, Science, Sports,
and Culture of Japan, No.\ 12047201, No.\ 13740138 and Priority Area
``Supersymmetry and Unified Theory of Elementary Particles'' (No.\ 
707).

\end{document}